\documentclass[doublecol]{epl2}

\title{Fermi surface topology and impurity-induced resonance state in BiS$_{2}$-based superconductors}
\shorttitle{Fermi surface topology and impurity-induced resonance state in BiS$_{2}$-based superconductors}

\author{Bin Liu\inst{1} \and Shiping Feng\inst{2}}
\shortauthor{Bin Liu and Shiping Feng}

\institute{\inst{1} Department of Physics, Beijing Jiaotong University, Beijing 100044, China\\
  \inst{2} Department of Physics, Beijing Normal University, Beijing 100875, China}

\pacs{74.20.Rp}{Pairing symmetries}
\pacs{74.62.En}{Effects of disorder}
\pacs{74.55.+v}{Tunneling phenomena}

\abstract{Within the two-orbital model for BiS$_{2}$-based superconductors, the effect from a single nonmagnetic impurity scattering on the superconducting-state is investigated in terms of the T-matrix approach. By considering three kinds of the typical Fermi surface topology which characterize the essential features of the doping dependence of the electronic structure in BiS$_{2}$-based superconductors, it is found that the impurity scattering on the superconducting-state with the conventional $s$-wave, the extended $s$-wave, the $\pm s$-wave, and the $d_{x^{2}-y^{2}}$-wave symmetries induces \textit{qualitatively} different resonance states due to the evolution of unique nodal structures on the Fermi surface with doping. These impurity-induced resonance states can be verified directly by the scanning tunneling microscopy experiments, and therefore they are proposed as a probe of the superconducting pairing symmetry in BiS$_{2}$-based superconductors.}

\begin{document}

\maketitle



The recent discovery of superconductivity in BiS$_{2}$-based layers materials, such as BiO$_{4}$S$_{3}$ and ReO$_{1-x}$F$_{x}$BiS$_{2}$ (Re=La, Nd, Ce, Pr) \cite{Mizuguchi,Mizuguchi1,Demura,Xing,Jha,Tan,Singh,Wan,Lei,Usui,Li}, has attracted considerable attention, since the physical properties in these unconventional superconductors mainly depend on the extent of the charge carrier doping, similar to the well-known cuprate-based and iron-based superconductors. For LaO$_{1-x}$F$_{x}$BiS$_{2}$, superconductivity \cite{Mizuguchi,Mizuguchi1} emerges when the doping concentration $x\geq 0.5$. However, NdO$_{1-x}$F$_{x}$BiS$_{2}$ becomes a superconductor only at the doping range \cite{Demura} of $0.1\leq x\leq 0.7$. On the other hand, the ration of the superconducting (SC) gap and SC transition temperature $T_{\rm c}$ observed from the tunneling spectra \cite{Li} is much larger than the one predicted by the BCS theory in the weak coupling regime, which therefore suggests the strong coupling superconductivity in BiS$_{2}$-based superconductors.

The band-structure calculations \cite{Usui} indicate that the relevant bands crossing the Fermi surface (FS) originate mainly from the Bi 6$p$ orbitals, where the FS topology remarkably changes with doping. In particular, the electron pockets located around the $[0,\pm\pi]$ or $[\pm\pi,0]$ $(K)$ point at the low doping regime evolve into two hole pockets located around the [0,0] ($\Gamma$) and $[\pi,\pi]$ $(M)$ points, respectively, with increasing doping concentration. However, the new electron pockets located around the $K$ point appear again at the high doping concentration $x\geq 0.5$, and therefore there is a coexistence of the electron pockets and hole pockets. Moreover, as a consequence of the unconventional superconductors, it has been argued that in analogy with the iron-based and cuprate-based superconductors \cite{Stewart,Martins}, the spin excitations with the short-range order may act like a bosonic glue to hold the electron pairs together, then these electron Cooper pairs condensation reveals the SC-state. In this case, the unexpected finding of superconductivity in BiS$_{2}$-based superconductors has raised the hope that it may help to solve the unusual physics in the iron-based and cuprate-based superconductors.

In a superconductor, the crucial requirement is to confirm the symmetry of the SC-state, since the understanding of the SC pairing symmetry will help to reveal the underlying SC mechanism. The SC pairing symmetry in BiS$_{2}$-based superconductors has been discussed recently within the weak-coupling approach \cite{Martins,Zhou}, where the possible SC pairing symmetries, such as the conventional $s$-wave, the extended $s$-wave, the $d_{x^{2}-y^{2}}$-wave as well as the spin triplet p-wave, have been proposed \cite{Martins,Zhou,Gao}. However, no general consensus has been reached so far due to the absence of direct experimental evidences. In this paper, we use local electronic structure around a single nonmagnetic impurity to probe the pairing symmetry in BiS$_{2}$-based superconductor, since such properties have proved to be successful in identifying pairing symmetries in other different classes of superconductors \cite{Zhu}. Our results show that the effect from a single nonmagnetic impurity scattering on the SC-state with the conventional $s$-wave, the extended $s$-wave, the $\pm s$-wave, and the $d_{x^{2}-y^{2}}$-wave symmetries may induce \textit{qualitatively} different resonance states as a result of unique nodal structures on different FS topology with doping. Since these impurity-induced resonance states can be verified directly by the scanning tunneling microscopy (STM) experiments, they are proposed as a test of the SC pairing symmetry in BiS$_{2}$-based superconductors.

\begin{figure}
\begin{center}
\includegraphics[width=0.8\linewidth]{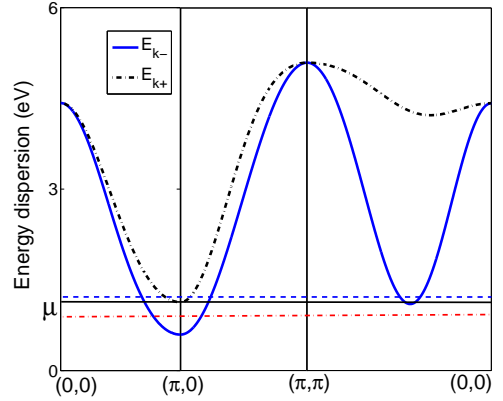}
\end{center}
\caption{(Color online) The energy dispersion along the high symmetric directions. The red dash-dotted, black solid, and blue dashed lines denote the Fermi energies at $x=0.25$, $x=0.45$, and $x=0.55$, respectively.
\label{fig1}}
\end{figure}

It has been shown from the first principles calculations that the essential physics of the doping dependence of the electronic structure in BiS$_{2}$-based superconductors is captured by a two-orbital model \cite{Usui},
\begin{eqnarray}\label{model}
H_{0}=\sum_{\bf k\sigma}\psi^{\dag}_{\bf k\sigma}\left (\matrix{\varepsilon^{X}_{\bf k}-\mu &\varepsilon^{XY}_{\bf k}
\cr \varepsilon^{XY}_{\bf k} &\varepsilon^{Y}_{\bf k}-\mu\cr}\right)\psi_{\bf k\sigma},
\end{eqnarray}
where $\psi^{\dag}_{\bf k\sigma}=(c^{\dag}_{X,\bf k\sigma}, c^{\dag}_{Y,\bf k\sigma})$ is the creation operator for spin $\sigma$ electrons in the orbitals $p_{X}$ and $p_{Y}$, while the matrix elements $\varepsilon^{X}_{\bf k}$ , $\varepsilon^{Y}_{\bf k}$, and $\varepsilon^{XY}_{\bf k}$ and their hopping parameters have been given in Ref. 11. The hybridized electron energy spectra of the Hamiltonian (\ref{model}) can be obtained straightforwardly as,
\begin{eqnarray}
E_{\bf k\pm}&=&\frac{\varepsilon^{X}_{\bf k}+\varepsilon^{Y}_{\bf k}}{2}\pm\sqrt{(\frac{\varepsilon^{X}_{\bf k}-\varepsilon^{Y}_{\bf k}}{2})^{2}+(\varepsilon^{XY}_{\bf k})^{2}},
\end{eqnarray}
with $E_{\bf k+}$ (blue solid line) and $E_{\bf k-}$ (black dash-dotted line) along the high symmetric directions are plotted in Fig. \ref{fig1}, where the red dash-dotted, black solid, and blue dashed lines denote the Fermi energy at the doping concentrations $x=0.25$, $x=0.45$, and $x=0.55$, respectively. It should be emphasized that although the Fermi energies only at three doping concentrations are shown in Fig. \ref{fig1}, they actually represent three kinds of the typical FS topology in BiS$_{2}$-based superconductors as mentioned above.

In spite of the unconventional SC mechanism, the bare Green's function of BiS$_{2}$-based superconductors in the SC-state can be expressed phenomenologically in the Nambu representation as,
\begin{eqnarray}
\hat{G}^{-1}_{0}({\bf k}, \omega)=\omega\hat{1}-\hat{H_{\bf k}},
\end{eqnarray}
where $\hat{1}$ is a unit matrix, while $\hat{H_{\bf k}}$ is defined as,
\begin{eqnarray}
\hat{H_{\bf k}}&=&(\frac{\tau_{0}+\tau_{z}}{2})\otimes(E_{\bf k+}\sigma_{z}+\Delta_{\bf k}\sigma_{x})\nonumber\\&+&(\frac{\tau_{0}-\tau_{z}}{2})\otimes(E_{\bf k-}\sigma_{z}+\Delta_{\bf k}\sigma_{x}).
\end{eqnarray}
with $\tau_{i}$ and $\sigma_{i}$ are the Pauli matrices, and $\tau_{i}\otimes\sigma_{i}$ denotes a direct product of the matrices operating on the four-dimensional space. The SC gap is described by $\Delta_{\bf k}$, with the conventional s-wave $\Delta_{\bf k}=\Delta_{0}$, the extended s-wave $\Delta_{\bf k}=\Delta_{0}[{\rm cos}k_{x}+{\rm cos} k_{y}] /2$, and the d-wave $\Delta_{\bf k}=\Delta_{0}[{\rm cos}k_{x}-{\rm cos}k_{y}]/2$. For the convenience of comparison, the same magnitude of the SC gap parameter $\Delta_{0}$ for all SC gaps with different symmetries has been used. In this case, the density of states (DOS) is obtained as,
\begin{eqnarray}\label{DOS}
\rho_{\rm DOS}(\omega)=-{1\over\pi}{1\over N}\sum_{\bf k}{\rm Im}{\rm Tr}\hat{G}_{0}({\bf k},\omega),
\end{eqnarray}
while the corresponding bare Green's function in real-space is obtained in terms of the Fourier transform as,
\begin{eqnarray}\label{bare-GF}
\hat{G}_{0}(i,j;\omega)=\frac{1}{N}\sum_{\bf k}e^{i\bf k\cdot\bf R_{ij}}\hat{G}_{0}({\bf k},\omega),
\end{eqnarray}
where $\bf R_{ij}=\bf R_{i}-\bf R_{j}$ with $\bf R_{i}$ is lattice vector, and N is the number of lattice sites. In the presence of a single nonmagnetic impurity with the scattering strength $U$ located at the origin ${\bf r}_{i}=0$, the bare Green's function (\ref{bare-GF}) is dressed via the impurity scattering, and then this dressed Green's function can be evaluated within the T-matrix approach as \cite{Zhu},
\begin{eqnarray}
\hat{G}(i,j;\omega)&=&\hat{G}_{0}(i,j;\omega)\nonumber\\&+&\hat{G}_{0}(i,0;\omega)\hat{T}(\omega)\hat{G}_{0}(0,j;\omega),
\end{eqnarray}
where $\hat{T}(\omega)=\hat{U}/[\hat{1}-\hat{G}_{0}(0,0;\omega)\hat{U}]$, and the impurity scattering potential $\hat{U}$ is simply given as $U_{ij}=U\delta_{ij}$ for $1\leq i\leq 2$, and $U_{ij}=-U\delta_{ij}$ for $3\leq i\leq 4$.

The local density of states (LDOS), which is proportional to the local differential tunneling conductance measured by STM experiments, now can be expressed as,
\begin{eqnarray}\label{LDOS}
\rho_{\rm LDOS}(i,\omega)=-\frac{1}{\pi}{\rm Im} {\rm Tr}\hat{G}(i,i;\omega).
\end{eqnarray}

\begin{figure}
\begin{center}
\includegraphics[width=0.8\linewidth]{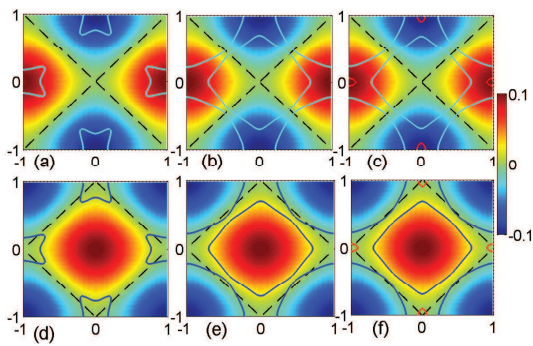}
\end{center}
\caption{(Color online) The Fermi surface topology in the superconducting-state with the $d_{x^{2}-y^{2}}$-wave symmetry (a-c) and extended $s$-wave symmetry (d-f) at $x$=0.25 (a and d), $x$=0.45 (b and e), and $x$=0.55 (c and f). The dashed lines denote the nodal lines.
\label{fig2}}
\end{figure}

In the absence of the impurity scattering, the FS topology in the SC-state with the $d_{x^{2}-y^{2}}$-wave symmetry (a-c) and extended $s$-wave symmetry (d-f) at $x$=0.25 (a and d), $x$=0.45 (b and e), and $x$=0.55 (c and f) is plotted in Fig. \ref{fig2}. It is shown clearly that at the low doping concentration $x$=0.25, only the electron pockets appear around the $M$ point. With increasing doping concentration, the FS topology changes significantly, and in particular, two hole pockets around the $\Gamma$ and $M$ points occur at $x$=0.45. However, at the high doping concentration $x$=0.55, besides the hole pockets, four electron pockets around the $M$ point emerge again. These results imply that the SC pairing symmetry in BiS$_{2}$-based superconductors might be quite different at different doping levels.
\begin{figure}
\begin{center}
\includegraphics[width=0.8\linewidth]{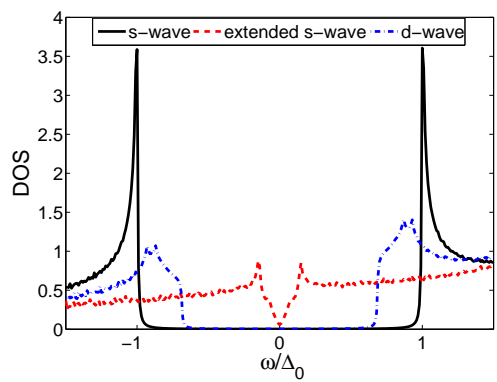}
\end{center}
\caption{(Color online) The energy dependence of the density of states at $x$=0.25 for different superconducting pairing symmetries.
\label{fig3}}
\end{figure}
\begin{figure}
\begin{center}
\includegraphics[width=0.8\linewidth]{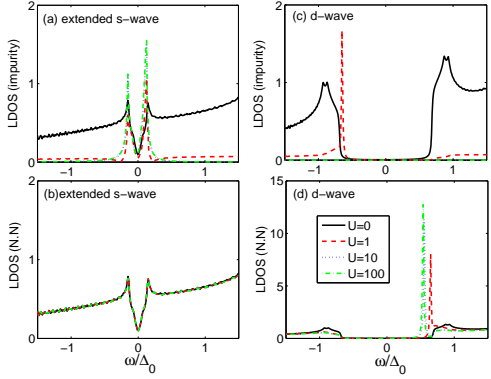}
\end{center}
\caption{(Color online) The energy dependence of the local density of states on the impurity site and its nearest-neighbor site for (a) and (b) the extended $s$-wave symmetry, and (c) and (d) the $d_{x^{2}-y^{2}}$-wave symmetry at $x$=0.25 with the impurity scattering strength $U$=0, 1, 10, 100 (in the unit of eV).
\label{fig4}}
\end{figure}

Now we turn to discuss the impurity-induced resonance state. Firstly, it should be noted that in the case of the conventional s-wave symmetry with a full SC gap, the robust two resonance peaks induced by a single nonmagnetic impurity scattering always locate at the gap edges $\pm\Delta_{0}$ regardless of the impurity scattering strength. These resonance states are called as Yu-Shiba-Rusinov states\cite{Yu}.

For a convenience in the following discussions, we plot DOS as a function of energy with the conventional s-wave symmetry (black solid line), the extended $s$-wave symmetry (red dashed line), and the $d_{x^{2}-y^{2}}$-wave symmetry (blue dash-dotted line) at $x=0.25$ in Fig. \ref{fig3}, where for the case of the d-wave symmetry, DOS displays a U-shaped behavior, which is similar to the case of the conventional s-wave symmetry, reflecting the gap structure of the $d_{x^{2}-y^{2}}$-wave case has no sign change within each electron pocket as shown in Fig. \ref{fig2}a. On the other hand, a V-shaped DOS appears in the case of the extended $s$-wave symmetry, which is the result of the nodal line crossing electron pocket as shown in Fig. \ref{fig2}d, however, the distance between two coherent peaks is much smaller than that in the $d_{x^{2}-y^{2}}$-wave case, reflecting a fact that there is no enough condensation energy in the SC-state with the extended $s$-wave symmetry.

In comparison with the case in the absence of the impurity shown in Fig. \ref{fig3}, we plot LDOS as a function of energy on the impurity site and its nearest-neighbor (NN) site for (a) and (b) the extended $s$-wave symmetry, and (c) and (d) the $d_{x^{2}-y^{2}}$-wave symmetry at $x$=0.25 with the impurity scattering strength $U$=0, 1, 10, 100 (in the unit of eV) in Fig. \ref{fig4}. We find that in the SC-state with the extended $s$-wave symmetry, two resonance peaks only located at the coherent gap edges irrespective of the impurity scattering strength appear on the impurity site, while they are absent from the NN site, indicating the strong localization of the nonmagnetic impurity scattering. However, the \textit{intragap} resonance state emerges in the SC-state with the $d_{x^{2}-y^{2}}$-wave symmetry on both impurity site and its NN site regardless of the U-shaped DOS in Fig. \ref{fig3}. In particular, this intragap resonance peak moves to the coherent gap edge with decreasing impurity scattering strength U. Moreover, it is most surprising that this intragap resonance peak is located far away from the Fermi energy even in the unitary scattering limit, which is in contrast to the case of the cuprate-based superconductors, where the impurity-induced resonance peak occurs near the Fermi energy \cite{Pan}. It should be mentioned that the impurity-induced \textit{intragap} resonance states reflect the SC pair breaking, while the resonance states located at the gap edges obey the Anderson's theorem \cite{Anderson}.

\begin{figure}
\begin{center}
\includegraphics[width=0.8\linewidth]{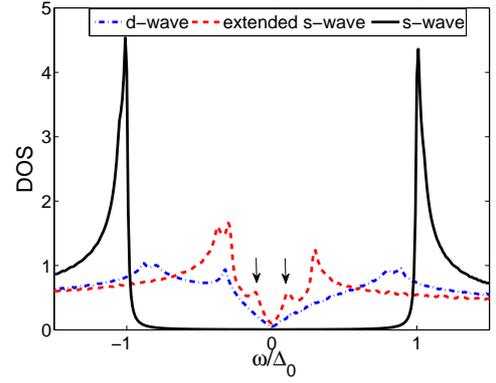}
\end{center}
\caption{(Color online) The energy dependence of the density of states at $x$=0.45 for different superconducting pairing symmetries.
\label{fig5}}
\end{figure}

\begin{figure}
\begin{center}
\includegraphics[width=0.8\linewidth]{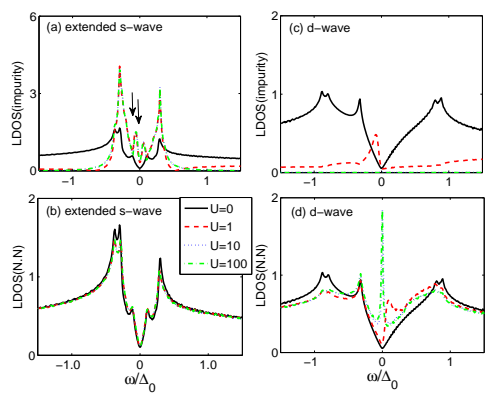}
\end{center}
\caption{(Color online) The energy dependence of the local density of states on the impurity site and its nearest-neighbor site for (a) and (b) the extended $s$-wave symmetry and (c) and (d) the $d_{x^{2}-y^{2}}$-wave symmetry at $x$=0.45 with the impurity scattering strength $U$=0, 1, 10, 100 (in the unit of eV).
\label{fig6}}
\end{figure}

For a better understanding of the impurity-induced resonance state at different doping levels, the energy dependence of DOS and LDOS on the impurity site and its NN site at $x$=0.45 are plotted in Fig. \ref{fig5} and Fig. \ref{fig6}, respectively. Comparing Fig. \ref{fig5} with Fig. \ref{fig3} for the same set of parameters except for $x$=0.45, we see clearly that for both extended $s$-wave and $d_{x^{2}-y^{2}}$-wave symmetries, a V-shaped DOS emerges at $x$=0.45, reflecting the appearance of the nodal structure in these SC pairing symmetries, however, the distance between two coherent peaks in DOS is still much smaller. Moreover, it is interesting that two new coherent peaks denoted by two arrows appear at lower energies for the extended $s$-wave symmetry, which originate from the sign change within each hole pocket as shown in Fig. \ref{fig2}b. In this case, as shown in Fig. \ref{fig6}a, besides two impurity-induced resonance peaks located at the large coherent gap edges, other two impurity-induced \textit{intragap} resonance peaks emerge within the new coherent gap, however, as shown in Fig. \ref{fig6}b, there is still no impurity resonance states on the NN site. On the other hand, for the d-wave symmetry, the impurity-induced \textit{intragap} resonance peak always survive as shown in Fig. \ref{fig6}c, and in particular, it shift to the Fermi energy with increase of the impurity scattering strength U as shown in Fig. \ref{fig6}d, which is very similar to the case in the cuprate-based superconductors.

\begin{figure}
\begin{center}
\includegraphics[width=0.8\linewidth]{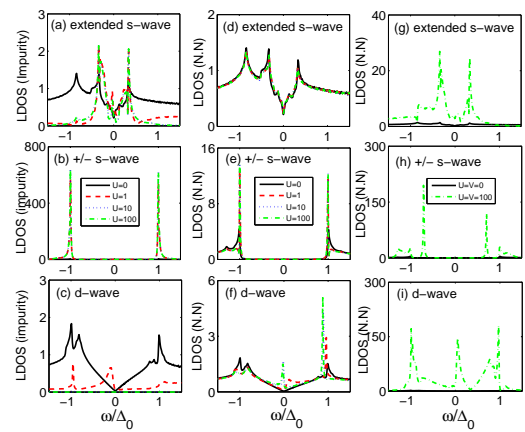}
\end{center}
\caption{(Color online) The local density of states as a function of energy on the impurity site and its nearest-neighbor site at $x=0.55$ for (a) and (d) the extended $s$-wave symmetry, (b) and (e) the $\pm s$-wave symmetry, and (c) and (f) the $d_{x^{2}-y^{2}}$-wave symmetry with the impurity scattering strength $U$=0, 1, 10, 100 (in the unit of eV). The local density of states as a function of energy on the nearest-neighbor site for (g) the extended $s$-wave symmetry, (h) the $\pm s$-wave symmetry, and (i) the $d_{x^{2}-y^{2}}$-wave symmetry with the additional interband impurity scattering under the strength V=U=100 eV.
\label{fig7}}
\end{figure}

Finally, we discuss the impurity-induced resonance state at the optimal doping concentration $x$=0.55. In Fig. \ref{fig7}, we plot LDOS as a function of energy on the impurity site and its NN site for (a) and (d) the extended $s$-wave symmetry and (c) and (f) the $d_{x^{2}-y^{2}}$-wave symmetry. For comparison, LDOS for the $\pm s$-wave symmetry, which may be valid for the iron-based superconductors \cite{Kuroki}, is also plotted in Fig. \ref{fig7}b and  Fig. \ref{fig7}e, where we assume $\Delta_{0}>0$ for hole pockets and $\Delta_{0} <0$ for electron pockets. Obviously, in the case of the extended $s$-wave symmetry, the feature of the impurity-induced resonance states is very similar to the case at $x$=0.45, and the impurity effect is robust localized on the impurity site. Moreover, the behavior of LDOS for the $\pm s$-wave symmetry, arising from two Yu-Shiba-Rusinov states at the coherent gap edges, is the same with the case of the conventional s-wave symmetry. However, four SC coherent peaks appear in the case of the $d_{x^{2}-y^{2}}$-wave symmetry, reflecting the two energy band features as shown in Fig. \ref{fig2}d. We have also noted that all coherent gap edges are very close to $\pm\Delta_{0}$. In this case, there are two kinds of the impurity-induced \textit{intragap} resonance peaks in Fig. \ref{fig7}f: one is located at lower energy and shifts to the Fermi energy with the increase of the impurity scattering strength U, while other occurs near the coherent gap edge even in the strong impurity scattering limit, and moves to the coherent gap edge with the decrease of U.

In particular, at the doping concentration $x$=0.55, the FS topology is enclosed by both electron pockets and hole pockets contributed from bands $E_{\bf k+}$ and $E_{\bf k-}$ respectively. Therefore, we have further made a calculation for LDOS by considering the additional interband scattering with the strength $V$, and the results are plotted in Fig. \ref{fig7}g-\ref{fig7}i. In comparison with the corresponding results in Fig. \ref{fig7}d-\ref{fig7}f in the absence of the interband scattering, it is found that the weight of the impurity-induced resonance peaks are strongly enhanced by the interband scattering for all SC pairing symmetries under consideration, however, for both extended $s$-wave and $\pm s$-wave symmetries, the new \textit{intragap} resonance states induced by the interband scattering appear at the NN site, while the mainly feature of LDOS in the absence of the interband scattering for the $d_{x^{2}-y^{2}}$-wave symmetry remains.


In conclusion, based on the two-orbital model for BiS$_{2}$-based superconductors, we have shown that the effect from a single nonmagnetic impurity scattering on the SC-state with the conventional $s$-wave, the extended $s$-wave, the $\pm s$-wave, and the $d_{x^{2}-y^{2}}$-wave symmetries can induce \textit{qualitatively} different resonance states. In particular, these impurity-induced resonance states are proposed as a probe of the SC pairing symmetry in BiS$_{2}$-based superconductors, since they can be verified directly by the STM experiments.

This work is supported by the National Natural Science Foundation of China (NSFC) under Grant No. 11104011, Doctoral Fund of Ministry of Education of China under Grant No.20110009120024, Beijing Higher Education Young Elite Teacher Project under Grant No. YETP0576, and Research Funds of Beijing Jiaotong University under Grant No. 2013JBM092. SF would like to thank the funds from the Ministry of Science and Technology of China under Grant Nos. 2011CB921700 and 2012CB821403, and NSFC under Grant Nos. 11074023 and 11274044.


\begin{thebibliography}{0}
\bibitem{Mizuguchi} Mizuguchi Y, Fujihisa H, Gotoh Y, Suzuki K, Usui H, Kuroki K, Demura S, Takano Y, Izawa H, and Miura O, Phys. Rev. B. {\bf 86}, 2012, R220510.
\bibitem{Mizuguchi1} Mizuguchi Y, Demura S, Deguchi K, Takano Y, Fujihisa H, Gotoh Y, Izawa H, and Miura O, J. Phys. Soc. Jpn. {\bf 81}, 2012, 114725.
\bibitem{Demura} Demura S, Mizuguchi Y, Deguchi K, Okazaki H, Hara H, Watanabe T, Denholme S J, Fujioka M, Ozaki T, Fujihisa H, Gotoh Y, Miura O, Yamaguchi T, Takeya H, and Takano Y, J. Phys. Soc. Jpn., {\bf 82}, 2013, 033708.
\bibitem{Xing} Xing J, Li S, Ding X, Yang H, and Wen H H, Phys. Rev. B. {\bf 86}, 2012, 214518.
\bibitem{Jha} Jha R, Kumar A, Singh S, and Awana V, J. Sup. and Novel. Mag. {\bf 26}, 2013, 499.
\bibitem{Tan} Tan S, Li L, Liu Y, Tong P, Zhao B, Lu W, and Sun Y, Physica C, {\bf 483}, 2012, 94.
\bibitem{Singh} Singh S, Kumar A, Gahtori B, Kirtan S, Sharma G, Patnaik S, and Awana V, J. Am. Chem. Soc. {\bf 134}, 2012, 16504.
\bibitem{Wan} Wan X, Ding H C, Savrasov S, and Duan C G, Phys. Rev. B. {\bf 87}, 2013, 115124.
\bibitem{Lei} Lei H, Wang K, Abeykoon M, Bozin E S, and Petrovic C, Inorg. Chem. {\bf 52}, 2013, 10685.
\bibitem{Li} Li S, Yang H, Tao J, Ding X, and Wen H H, Science China Physics, Mechanics and Astronomy {\bf 56}, 2013, 2019.
\bibitem{Usui} Usui H, Suzuki K, and Kuroki K, Phys. Rev. B. {\bf 86}, 2012, R220501.
\bibitem{Stewart} Stewart G R, Rev. Mod. Phys. {\bf 83}, 2011, 1589.
\bibitem{Martins} Martins G B, Moreo A, and Dagotto E, Phys. Rev. B. {\bf 87}, 2013, 081102.
\bibitem{Zhou} Zhou T, and Wang Z D, J. Sup. and Novel. Mag. {\bf 26}, 2013, 2735.
\bibitem{Gao} Gao Y, 2013, arXiv:1304.2102.
\bibitem{Zhu} Balatsky A V, Vekhter I, and Zhu J -X, Rev. Mod. Phys. {\bf 78}, 2006, 373.
\bibitem{Yu} Yu L, Acta Phys. Sin. {\bf 21}, 1965, 75; Shiba H, Prog. Theor. Phys. {\bf 40}, 1968, 435; Rusinov A I, Sov. Phys. JETP {\bf 29}, 1969, 1101.
\bibitem{Pan} Pan S H, Hudson E W, Lang K M, Eisaki H, Uchida S, and Davis J C, Nature {\bf 413}, 2001, 282.
\bibitem{Anderson} Anderson P W, J. Phys. Chem. Solids {\bf 11}, 1959, 26.
\bibitem{Kuroki} Kuroki K, Onari S, Arita R, Usui H, Tanaka Y, Kontani H, and Aoki H, Phys. Rev. Lett. {\bf 101}, 2008 087004; Kuroki K, Usui H, Onari S, Arita R, and Aoki H, Phys. Rev. B {\bf 79}, 2009 224511; Mazin I I, Singh D J, Johannes M D, and Du M H, Phys. Rev. Lett. 101, 057003 (2008).


\end{thebibliography}
\end{document}